# An Improved Timestamp-Based Password Authentication Scheme Using Smart Cards

Al-Sakib Khan Pathan and Choong Seon Hong
Department of Computer Engineering, Kyung Hee University, Korea
spathan@networking.khu.ac.kr and cshong@khu.ac.kr

*Abstract* — With the recent proliferation of distributed systems and networking, remote authentication has become a crucial task in many networking applications. Various schemes have been proposed so far for the two-party remote authentication; however, some of them have been proved to be insecure. In this paper, we propose an efficient timestamp-based password authentication scheme using smart cards. We show various types of forgery attacks against a previously proposed timestamp-based password authentication scheme and improve that scheme to ensure robust security for the remote authentication process, keeping all the advantages that were present in that scheme. Our scheme successfully defends the attacks that could be launched against other related previous schemes. We present a detailed cryptanalysis of previously proposed Shen et. al's scheme and an analysis of the improved scheme to show its improvements and efficiency.

## 1. Introduction

Remote Authentication is an important task in many networking applications. The legitimate users might have to login to the system remotely. One of the major hurdles in remote authentication process is ensuring robust security while using an insecure channel at the time of communications between the user and the authentication server. Various works addressed this issue from different perspectives which include password-table driven schemes, id-based schemes, timestamp based schemes, nonce-based schemes etc. Nonetheless, many of the works which dealt with this issue are able to provide only the unilateral authentication where the server is considered to be completely secured and only the legitimacy of the user(s) could be verified. As the adversaries could intercept the login requests from the users and might pretend to be the legitimate server(s) to the users, there must be some sorts of mechanisms to ensure authentication in both directions, which is termed as *bilateral or mutual authentication*. One of the mutual authentication (or, verification) schemes proposed earlier is Shen et. al. scheme [1] which is basically based on the basic timestamp-based scheme proposed by Yang and Shieh [2]. In this paper, we propose an improved timestamp-based mutual authentication scheme which is adapted from Shen et. al. scheme. We eliminate the weaknesses and vulnerabilities of Shen et. al. scheme to provide enhanced security for remote authentication process using smart cards. We also show a new type of attack and some already identified attacks and weaknesses of Shen et. al. scheme. In addition to ensuring robust security, our scheme allows mutual verification among the participating entities and password renewal by the users.

The rest of the paper is organized as follows: Section 2 states the related works, Section 3 reviews Shen et. al. scheme, Section 4 presents a review of various types of attacks and weaknesses of Shen et. al. scheme, Section 5 contains the details of our proposed scheme, Section 6 presents the analysis of our scheme and finally Section 7 concludes the paper.

## 2. Related Works

In 1999, Yang and Shieh [2] proposed two password authentication schemes with smart cards one of which was the timestamp-based password authentication scheme. In 2002, Chan and Cheng [3] showed that [2] is vulnerable to forged login attack and an adversary could be able to impersonate as a legal user to pass the system authentication. Fan et. al. [4] presented a cryptanalysis of [2] and showed another type of attack different than that is in [3] and also proposed an enhanced scheme which could withstand Chan-Cheng attack and their demonstrated attack. But, later [5] showed that, Fan et. al. scheme was still insecure and vulnerable to forged login attack. Again, [6] showed two other attacks on Fan et. al.'s enhanced scheme. Shen et. al. [1] came up with one enhanced scheme based on [2] which they claimed to be efficient enough to protect the authentication process from forged login or forged server attacks. Unfortunately, later [7], [8] and [9] showed that the improved scheme proposed by Shen et. al. was still vulnerable to the forgery attacks. Wang and Li [10] proposed another improved scheme based on Yang-Shieh scheme [2] assuming that the remote host had an extra storage for storing certain information.

Our proposed scheme is different than all of the mentioned schemes and overcomes the drawbacks of these existing schemes. Our Analysis shows that, it could well-defend all the attacks that are presented by these previous works. Moreover, if required, our scheme keeps almost all the advantages that were present in the other schemes.

## 3. Review of Shen Et. Al.'s Scheme

In this section we briefly review the Shen et. al. scheme [1]. First we note down the terms and preliminaries used throughout the rest of the paper.

This work was supported by the MIC and ITRC projects.

## 3.1 Basic Terms

$U_i$ – The $i$th user seeking for authentication by the server
KIC – The Key Information Center which is responsible for generating key information, issuing smart cards to new users and serving password-changing requests for registered users
$ID_i$ – The identity of the user $U_i$
$PW_i$ – The password chosen by the user $U_i$
$CID_i$ – The identity of the smart card associated to the user $U_i$
$f(.)$ – A one-way function

## 3.2 Shen et. al.'s Timestamp-Based Scheme

As this scheme is a modified and enhanced version of Yang-Shieh scheme [2], like [2] it also contains three phases for the authentication process: registration phase, login phase and authentication phase.

**Registration Phase.** In this phase, the KIC sets up the authentication system and issues smart cards to $U_i$ who requests for registration. It is assumed that, this phase occurs over a secure channel. The steps that the KIC follows in this phase are:
1. $U_i$ securely submits $ID_i$ and $PW_i$ to the KIC.
2. Two large prime numbers $p$ and $q$ are generated, and let $n = p \cdot q$
3. A prime number $e$ and an integer $d$ are chosen which satisfy, $e \cdot d \equiv 1 \mod (p-1)(q-1)$, where $e$ is the public key of the KIC that should be published and $d$ is the secret key that is kept secret.
4. An integer $g$ is found which is a primitive element in both $GF(p)$ and $GF(q)$, where $g$ is the public information of the KIC.
5. $S_i = ID_i^d \mod n$ is computed as $U_i$'s secret information.
6. $h_i$ for $U_i$ is computed such that, $h_i = g^{PW_i \cdot d} \mod n$.
7. $CID_i$ is computed as, $CID_i = f(ID_i \oplus d)$, where $\oplus$ stands for an exclusive operation.
8. Then the information $n$, $e$, $g$, $ID_i$, $CID_i$, $S_i$, $h_i$ and $f(.)$ are written into the smart card's memory and the card is issued to $U_i$

**Login Phase.** When $U_i$ needs to login to the system, the smart card should be attached to the login device and $ID_i$ and $PW_i$ need to be keyed in. After that, the smart cards performs the following operations:
1. Generates a random number $r_i$ and computes $X_i$ and $Y_i$ as follows:

$$X_i = g^{r_i \cdot PW_i} \mod n$$
$$Y_i = S_i \cdot h_i^{r_i \cdot f(CID_i, T)} \mod n$$

Here, T is the current timestamp.
2. Sends the login request message, $M = \{ID_i, CID_i, X_i, Y_i, n, e, g, T\}$ to the remote server.

**Verification Phase.** In this phase, the system or the server determines the validity of the received login request message and decides whether to accept the access of the user or not. So, after the server has received the message $M$, it carries out the following steps:
1. Checks the validity of $ID_i$. If the format of the $ID_i$ is incorrect, the server rejects the request.
2. Checks whether $CID_i' = CID_i$ holds or not, where, $CID_i' = f(ID_i \oplus d)$. If the result is positive, the following steps are performed otherwise the request is rejected.
3. Checks whether the condition (T'-T) $\leq \Delta$T holds or not, where T' is the timestamp of receiving the login request message and $\Delta$T is the legitimate time interval allowed for the transmission delay. If negative rejects the request.
4. Checks the equation, $Y_i^e = ID_i \cdot X_i^{f(CID_i, T)} \mod n$. If it holds, then the remote server accepts the login request and gives access to the $U_i$
5. Now, it computes $R = (f(CID_i, T''))^d \mod n$ where, T'' is the current timestamp and returns $M' = \{R, T''\}$ to the user $U_i$

When the user receives the message $M'$, the verification of the server by $U_i$ is done as follows:
1. Checks the time valid interval, (T'''-T'') $\leq \Delta$T, where T''' is the timestamp of receiving the message $M'$. If it is positive, it goes forward otherwise, rejects the server message.
2. Calculates $R' = R^e \mod n = (f(CID_i, T'')^d)^e = f(CID_i, T'')$. If the condition, $R' = f(CID_i, T'')$ does not hold, then the remote server is rejected, otherwise the mutual verification is succeeded.

## 4. Cryptanalysis of Shen et. al.'s Scheme

Shen et. al.'s scheme [1] is vulnerable to forgery attacks in various ways. In this section we review some of the attacks and weaknesses of [1] to better understand how the countermeasures could be developed to ensure robust security for the authentication process.

### 4.1 Attack Based on [9] and [6]

As the attacker could intercept the login request message $M = \{ID_i, CID_i, X_i, Y_i, n, e, g, T\}$ it can get the valid values of $ID_i$ and $CID_i$. Using these values it could launch the impersonation attack as follows:
1. Let, $a = f(CID_i, T_c)$ where $T_c$ is the current timestamp. Use the Extended Euclidean algorithm to compute $gcd(e, a) = 1$. Let, u and v be the coefficients computed by the Extended Euclidean algorithm such that, $e \cdot u - a \cdot v = 1$
2. Compute $X_f = ID_i^v \mod n$
3. Compute $Y_f = ID_i^u \mod n$
4. Send the forged login request message $M_f = \{ID_i, CID_i, X_f, Y_f, n, e, g, T_c\}$ and this request will eventually pass the authentication phase as,

$$Y_f^e = ID_i^{e \cdot u} \mod n$$
$$= ID_i^{1+a \cdot v} \mod n$$
$$= ID_i \cdot ID_i^{a \cdot v} \mod n$$
$$= ID_i \cdot (X_f)^a \mod n$$
$$= ID_i \cdot (X_f)^{f(CID_i, T_c)} \mod n$$

In fact, this attack could be extended for $gcd(e,a)=2,3,\ldots$ instead of only $gcd(e, a) = 1$.

### 4.2 Yang et. al.'s Attack

In this attack the attacker intercepts the message, $M = \{ID_i, CID_i, X_i, Y_i, n, e, g, T\}$ and then:
1. Finds a value $w$ such that it satisfies, $w \cdot f(CID_i, T_f) = f(CID_i, T)$, where $T_f$ denotes the attacker's attack launching time.
2. Computes the equation, $X_f = X_i^w = g^{r_i \cdot PW_i \cdot w} \mod n$
3. Now, the attacker constructs the forged login request message as, $M_f = \{ID_i, CID_i, X_f, Y_i, n, e, g, T_f\}$
This forged message eventually passes the authentication phase of [1] because:

$$Y_i^e = (S_i \cdot h_i^{r_i \cdot f(CID_i, T)})^e \mod n$$
$$= (ID_i^d \cdot g^{PW_i \cdot d \cdot r_i \cdot f(CID_i, T)})^e \mod n$$
$$= ID_i \cdot g^{PW_i \cdot r_i \cdot f(CID_i, T)} \mod n$$

and,

$$ID_i \cdot X_f^{f(CID_i, T_f)} \mod n = ID_i \cdot g^{r_i \cdot PW_i \cdot w \cdot f(CID_i, T_f)} \mod n$$
$$= ID_i \cdot g^{PW_i \cdot r_i \cdot f(CID_i, T)} \mod n$$

### 4.3 Impersonation Attack Based on [8]

An attacker can impersonate a legitimate user $U_i$, with identity $ID_i$, following the procedure:
1. Intercepts the login request message $M = \{ID_i, CID_i, X_i, Y_i, n, e, g, T\}$.
2. Computes, $ID_f = ID_i^{-1} \mod n$
3. Now, the attacker submits the identity $ID_f$ and a random value as his password to the KIC to obtain a valid smart card with information $\{n, e, g, ID_f, CID_f, S_k, h_k \text{ and } f(.)\}$.
4. Since, in the registration phase, $S_i = ID_i^d \mod n$ and here, $S_k = ID_f^d \mod n = ID_i^{-d} \mod n$, the attacker can compute $S_i$ as, $S_i = S_k^{-1} \mod n$
5. Chooses a random integer $y$.
6. Sets, $X_f = y^e \mod n$ and $Y_f = S_i \cdot y^{f(CID_i, T_f)} \mod n$ where $T_f$ is the timestamp for the login request from the attacker and sends the forged login message, $M_f = \{ID_i, CID_i, X_f, Y_f, n, e, g, T_f\}$. The request is validated as the login request from the user $U_i$ because,

$$Y_f^e = (S_i \cdot y^{f(CID_i, T_f)})^e \mod n$$
$$= ID_i^{ed} \cdot y^{f(CID_i, T_f) \cdot e} \mod n$$
$$= ID_i \cdot (X_f)^{f(CID_i, T_f)} \mod n$$

### 4.4 Another Type of Forgery Attack

The attacker can get the values of $ID_i$ and $CID_i$ from the login request message from the valid user, and $CID_i = f(ID_i \oplus d)$ is a fixed value for a particular login request from a user. The attacker could launch an attack using the following steps:
1. Let, $a = f(CID_i, T_f)$ where $T_f$ is the attacker's login timestamp. The attacker finds a value $b$ such that, $a \cdot b \equiv 1 \mod n$
2. It chooses a random integer $k$ and computes, $Y_f = k^{f(CID_i, T_f)} \mod n$ and sets $X_f = ID_i^{-b} \cdot k^e \mod n$.

3. Sends the forged login request message, $M_f = \{ID_i, CID_i, X_f, Y_f, n, e, g, T_f\}$
4. The attacker could pass the first phase of the authentication phase as,

$$Y_f^e = k^{f(CID_i, T_f) \cdot e} \mod n$$

and,

$$ID_i \cdot X_f^{f(CID_i, T_f)} \mod n = (ID_i \cdot ID_i^{-b} \cdot k^e)^{f(CID_i, T_f)} \mod n$$
$$= ID_i \cdot ID_i^{-1} \cdot k^{f(CID_i, T_f) \cdot e} \mod n$$
$$= k^{f(CID_i, T_f) \cdot e} \mod n$$

## 5. Our Improved Scheme

Like the scheme [1] our improved scheme also has three distinct but interrelated phases, registration phase, login phase and mutual authentication phase. We keep the registration phase same as [1] and improve the other phases to surmount the drawbacks mentioned earlier. So, after the registration phase is complete, $U_i$ gets the information $n, e, g, ID_i, CID_i, S_i, h_i$ and $f(.)$ written in the memory of the smart card.

**Login Phase.** In the login phase, $U_i$ attaches the smart card with the reader device and keys in his $ID_i$ and $PW_i$. Then the smart card performs the following operations:
1. Generates a random number $r_i$ and computes $X_i$ and $Y_i$ as follows:

$$X_i = g^{r_i \cdot PW_i} \mod n$$
$$Y_i = S_i \cdot h_i^{r_i \cdot f(CID_i, T)} \mod n$$
$$Z_i = X_i \oplus CID_i \oplus f(CID_i, Y_i)$$

Here, T is the current timestamp.
2. Sends the login request message, $M = \{ID_i, Y_i, Z_i, n, e, g, T\}$

**Mutual Authentication Phase.** When the server gets the login request message, it performs the operations:
1. Checks the validity of $ID_i$. If the format of the $ID_i$ is incorrect, the server rejects the request.
2. Checks whether the condition $(T'-T) \leq \Delta T$ holds or not, where $T'$ is the timestamp of receiving the login request message and $\Delta T$ is the legitimate time interval allowed for the transmission delay. If negative rejects the request.
3. Computes, $CID_i' = f(ID_i \oplus d)$ and $val = f(CID_i', Y_i)$. Then, computes $Z_i \oplus CID_i' \oplus val$ which should generate the value of $X_i$ as $CID_i' = f(ID_i \oplus d) = CID_i$ for the legitimate users.
4. Checks the equation, $Y_i^e = ID_i \cdot X_i^{f(CID_i, T)} \mod n$. If it holds, then the remote server accepts the login request and gives access to the $U_i$, otherwise rejects the request.
5. Once, the user $U_i$ is authenticated by the server, to provide mutual authentication, the server now computes, $R = (f(CID_i', T''))^d \mod n$ where, $T''$ is the current timestamp and returns $M' = \{R, T''\}$ to the user $U_i$
After receiving the message $M'$ the user $U_i$, checks it as follows:
1. Checks the time valid interval, $(T'''-T'') \leq \Delta T$, where $T'''$ is the timestamp of receiving the message $M'$. If it is positive, it goes forward otherwise, rejects the server message.

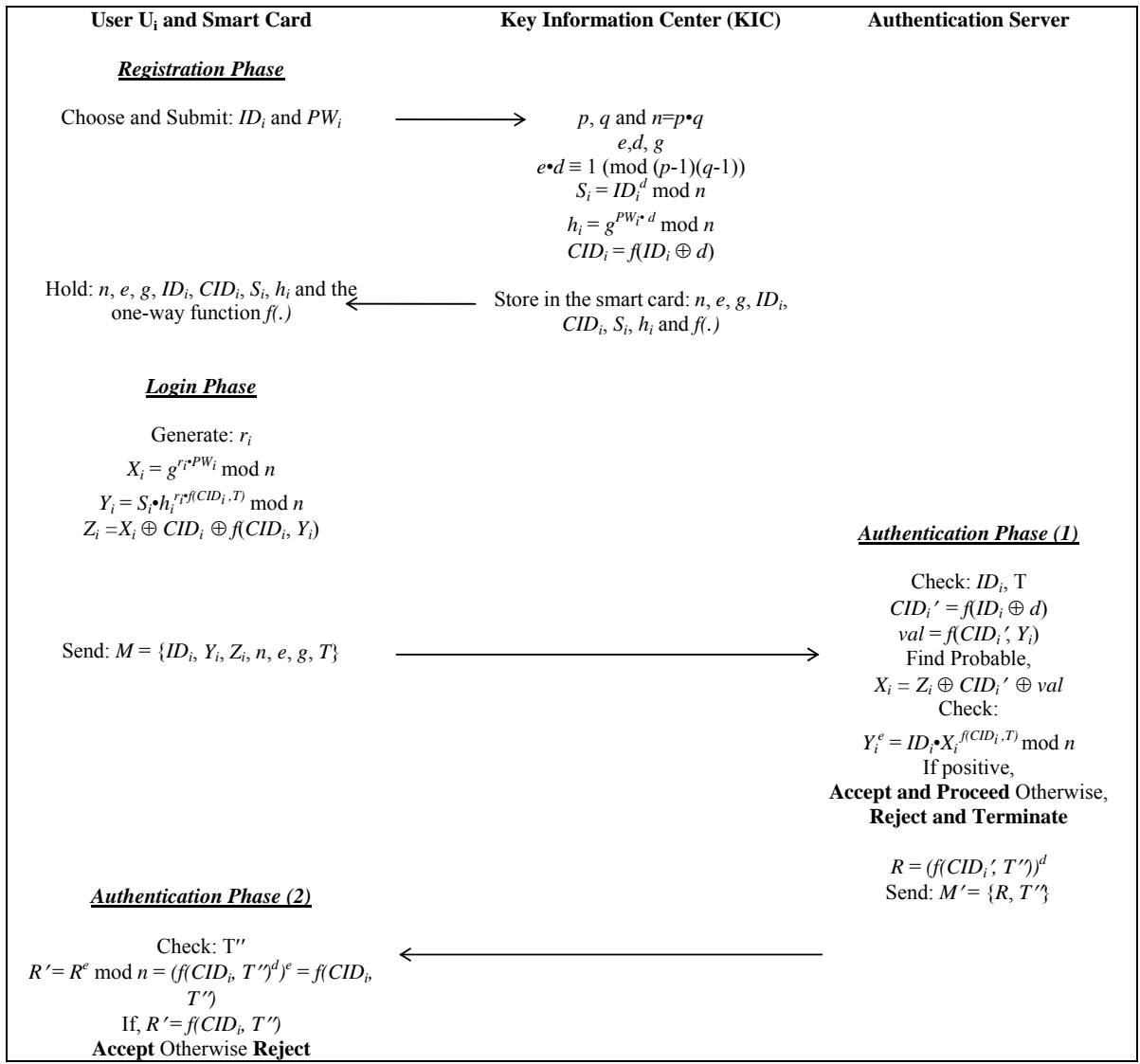

**Figure 1. Our Mutual Authnetication Scheme**

2. Calculates $R' = R^e \bmod n = (f(CID_i, T'')^d)^e = f(CID_i, T'')$. If the condition, $R' = f(CID_i, T'')$ does not hold, then the remote server is rejected, otherwise the mutual authentication is succeeded.

**Password Renewal.** If $U_i$ needs to change his password, he has to go through the registration phase where he submits his identity and the new password and accordingly the KIC performs the steps 5 to 8 for generating the information for that particular user.

Our scheme is shown at a glance in Figure 1.

## 6. Analysis of Our Scheme

In this section, we analyze our scheme. First, we discuss the mode of operations and the structure of our scheme. Then, considering the cryptanalysis of Shen et. al.'s scheme, we show how our scheme could resist the known attacks.

We know that, the registration phase sets the base of the authentication process. After the KIC writes the required information into the smart card's memory, the smart card works for the user. When any legitimate user attaches the card with the remote device and needs to login to the system, his ID and password are used to generate the cryptic information for starting the mutual authentication process. Three parameters are generated in the login phase in our scheme; $X_i$, $Y_i$ and $Z_i$. The values of $X_i$ and $Y_i$ are different for each individual user and depend on the values of $ID_i$, $CID_i$ and password for a particular user. The random value $r_i$ randomizes the outputs of these parameters from session to session for a particular user so that the outputs could not be same for each new login attempt from the user. The third parameter $Z_i$ is basically used to hide the value of $X_i$ and $CID_i$. As $CID_i$ is fixed for a particular user, if it is sent in plain format, the attacker could employ some other techniques to deduce some important information and thus could do harm to the legitimate user. In fact, we have shown that some of the attacks on [1] could be launched because of snatching the value of $CID_i$ (of a valid user) by the adversaries.

While sending the login request message in our scheme, total number of parameters used is seven, instead of eight required for Shen et. al.'s scheme. This obviously reduces the message size if all the other parameters are considered to be of the same sizes as in [1]. We use the login request message, M = {$ID_i$, $Y_i$, $Z_i$, $n$, $e$, $g$, $T$}. By examining the structure of the login request message, it is evident that, there is little information available for an adversary that could be useful for launching any of the previously presented attacks. As $Z_i$ hides the value of $CID_i$ by mingling (i.e. with XOR Operations) with $X_i$ and $f(CID_i, Y_i)$, there is no possible way for an adversary to find out the proper value of $CID_i$ for a legitimate user. The adversary can at best get the value of $Y_i$, from the login request message, but as $X_i$ and $CID_i$ are not known, it cannot decipher any of the actual values from the available information.

In our authentication phase, we have mainly two sub-phases. In the first sub-phase, the server verifies the authenticity of the user whereas in the second sub-phase, the user verifies the authenticity of the server message. At the time of user verification, the server first checks the ID and time constraints, then calculating $Z_i \oplus CID_i' \oplus val$ must generate the value of $X_i$ for a legitimate user because:

$Z_i \oplus CID_i' \oplus val$
$= X_i \oplus CID_i \oplus f(CID_i, Y_i) \oplus CID_i' \oplus val$
$= X_i \oplus f(CID_i, Y_i) \oplus val$
$= X_i \oplus f(CID_i, Y_i) \oplus f(CID_i', Y_i) = X_i$

Calculated $CID_i' = f(ID_i \oplus d) = CID_i$, must be hold for a legitimate user. Note that, as the value of $d$ is not public, this value could not be calculated by anyone except the legitimate server. After finding the possible value of $X_i$, we check the condition, $Y_i^e = ID_i \cdot X_i^{f(CID_i, T)} \mod n$ and validity of this ensures the authenticity of the user requesting for access.

To start the second sub-phase of authentication, after getting the message $M'$ from the user, the server checks the condition, $R' = f(CID_i, T'')$ which must hold for legal users as according to our assumption, $e \cdot d \equiv 1 \mod (p-1)(q-1)$.

Now, considering some of the common attacks, we show how our scheme could perform well.

*Replay Attack*: A replay attack is a form of network attack in which an attacker maliciously or fraudulently does the repeated or delayed transmission of a valid data [11], [12]. This is carried out either by the originator or by an adversary who intercepts the data and retransmits it, possibly as a part of a masquerade attack. In our scheme, replay attack is not possible as repetition of an old login request message will be detected by the server in the step 2 of the mutual authentication phase.

*Forged Login or Forged Server Attack*. Most of the attacks and weaknesses for Shen et. al. [1] scheme are found as the attackers could get the values of $ID_i$, $CID_i$, $X_i$, $Y_i$ and $T$ by intercepting the login request message. In our scheme, the values of $X_i$ and $CID_i$ are kept secret at the time of communication over insecure channel and are not available to the eavesdropper to use these for replacing them with other values. In fact, forging or replacing other values will be detected in the mutual authentication phase and eventually the request will be rejected. From the server side, the message $R$ could not be generated by the attacker as $d$ is not public and is kept secret only by the server. Even in this case, time stamping eliminates the chance of any sort of replay attack using the server message.

*Impersonation*: In our scheme, there is no way that an attacker could carry out an impersonation attack. Considering the impersonation attacks mentioned in section 4.1, 4.2, and 4.4, we could infer that, these attacks are not applicable against our scheme. Let us consider the impersonation attack presented in section 4.3. From the attacker's side:

1. It Intercepts the login request message. In our scheme it is, M = {$ID_i$, $Y_i$, $Z_i$, $n$, $e$, $g$, $T$}.
2. Computes, $ID_f = ID_i^{-1} \mod n$
3. Now, the attacker submits the identity $ID_f$ and a random value as his password to the KIC to obtain a valid smart card with information {$n$, $e$, $g$, $ID_f$, $CID_f$, $S_k$, $h_k$ and $f(.)$}.
4. Since, in the registration phase, $S_i = ID_i^d \mod n$ and here, $S_k = ID_f^d \mod n = ID_i^{-d} \mod n$, the attacker can compute $S_i$ as, $S_i = S_k^{-1} \mod n$
5. Chooses a random integer $y$.
6. Now, this step could not be performed as we do not expose the value of the parameter $X_i$. We have shown earlier that, this value is concealed in a way that could not be deduced by the adversary. At this point, the adversary could at best replace the value of $Y_i$ with, $Y_f = S_i \cdot y^{f(CID_i, T_f)} \mod n$ but, that simply does not make any sense and even the first phase of authentication could not be passed in any way.

So, the adversary cannot be able to impersonate as the legal user $U_i$. The attacker is allowed even to replace the value of $Z_i$, but without the proper values, any of such attempts can never pass the authentication phase.

## 7. Conclusions

In this paper, we have shown the weaknesses and different types of attacks on Shen et. al. scheme including a new type of attack. We have presented our improved scheme which could successfully defend all sorts of attacks mentioned earlier. We have presented in the related works section that, the other schemes in this area are more or less vulnerable to the attacks that are mentioned in this paper. Our scheme ensures robust security at the time of communication over the insecure channel and keeps all the other advantages that were present in the previous scheme.

REFERENCES

[1] Shen, J.-J., Lin, C.-W., and Hwang, M.-S., "Security Enhancement for the Timestamp-Based Password Authentication Schemes using Smart Cards", Computers & Security, Vol. 22, No 7, Elsevier, 2003, pp. 591–595.

[2] Yang, W.-H. and Shieh, S.-P., "Password Authentication Schemes with Smart Cards", Computers & Security, Vol. 18, No. 8, Elsevier, 1999, pp. 727-733.
[3] Chan, C.-K. and Cheng, L. M., "Cryptanalysis of a Timestamp-Based Password Authentication Scheme", Computers & Security, Vol. 21, No. 1, Elsevier, 2002, pp. 74-76.
[4] Fan, L., Li, J.-H., and Zhu, H.-W., "An Enhancement of Timestamp-Based Password Authentication Scheme", Computers & Security, Vol. 21, No. 7, Elsevier, 2002, pp. 665-667.
[5] Wang, B., Li, J.-H., and Tong, Z.-P., "Cryptanalysis of an Enhanced Timestamp-Based Password Authentication Scheme", Computers & Security, Vol. 22, No. 7, Elsevier, 2003, pp. 643-645.
[6] Chen, K.-F. and Zhong, S., "Attacks on the (Enhanced) Yang-Shieh Authentication", Computers & Security, Vol. 22, No. 8, Elsevier, 2003), pp. 725-727.
[7] Yang, C.-C., Yang, H.-W., and Wang, R. C., "Cryptanalysis of Security Enhancement for the Timestamp-Based Password Authentication Scheme using Smart Cards", IEEE Transactions on Consumer Electronics, Vol. 50, No. 2, 2004, pp. 578-579.
[8] Yang, L. and Chen, K., "Cryptanalysis of a Timestamp-Based Password Authentication Scheme", (2004) available at: http://eprint.iacr.org/2004/040.pdf
[9] Sun, H.-M. and Yeh, H.-T., "Further Cryptanalysis of a Password Authentication Scheme with Smart Cards", IEICE Transactions on Communications, Vol. E86-B, No. 4, 2003, pp. 1412-1415.
[10] Wang, Y. and Li, J., "Security Improvement on a Timestamp-Based Password Authentication Scheme", IEEE Transactions on Consumer Electronics, Vol. 50, No. 2, 2004, pp. 580-582
[11] Gong, L., "A Security Risk of Depending on Synchronized Clocks", ACM SIGOPS Operating Systems Review, Volume 26, Issue 1, January 1992, pp. 49 – 53.
[12] Syverson, P., "A Taxonomy of Replay Attacks", Proc. Computer Security Foundations Workshop VII, 1994, CSFW 7, 14-16 June, 1994, pp. 187 – 191.